# Predicting the Volumes of Crystals

Iek-Heng Chu,[†,a] Sayan Roychowdhury,[†,b] Daehui Han,[a] Anubhav Jain,[*,b] and Shyue Ping Ong[*,a]

[a] Department of NanoEngineering, University of California San Diego, 9500 Gilman Drive # 0448, La Jolla, CA 92093, USA.

[b]Lawrence Berkeley National Lab, 1 Cyclotron Road, Berkeley, CA, USA

† These authors contributed equally to this work.

Email: ajain@lbl.gov; ongsp@eng.ucsd.edu

# Abstract

New crystal structures are frequently derived by performing ionic substitutions on known crystal structures. These derived structures are then used in further experimental analysis, or as the initial guess for structural optimization in electronic structure calculations, both of which usually require a reasonable guess of the lattice parameters. In this work, we propose two lattice prediction schemes to improve the initial guess of a candidate crystal structure. The first scheme relies on a one-to-one mapping of species in the candidate crystal structure to a known crystal structure, while the second scheme relies on data-mined minimum atom pair distances to predict the crystal volume of the candidate crystal structure and does not require a reference structure. We demonstrate that the two schemes can effectively predict the volumes within mean absolute errors (MAE) as low as 3.8% and 8.2%. We also discuss the various factors that may impact the performance of the schemes. Implementations for both schemes are available in the open-source pymatgen software.

## 1. Introduction

To generate new materials that may potentially possess superior properties, a common strategy both experimentally and computationally is to perform partial or complete substitution of various species in a known crystal. The selection of substituents can be made either based on chemical intuition or by using quantitative data-mined substitution probabilities.[1] The derived candidates are then used for further experimental analysis (e.g., in the refinement of X-ray diffraction patterns), or as an initial guess to electronic structure calculations to determine its phase stability[2] and other application-specific properties,[3–7] for example, for energy storage,[4,8–11] solid-state lighting,[12] thermoelectrics,[13,14] catalysis,[15] etc.[16,17] In these analyses, a reasonable guess of the initial lattice parameters is necessary. For instance, the first step in the computational evaluation of any new candidate crystal involves the optimization of the lattice parameters and atomic positions to obtain the equilibrium geometry, and the closer the initially supplied lattice parameters and atomic positions are to the final equilibrium structure, the more likely the structure will converge at a reasonable speed.

For ionic-substitution-derived candidates, one often sets the initial lattice parameters and atomic positions to be identical to those of the parent structure. In cases where there are substantial size differences between the substituent and original atoms (e.g., for anion substitutions), this suboptimal guess can lead to large errors in structure refinement, as well as slow, or even failures in, convergence. As another use case of lattice scaling, many data mining descriptors, e.g., density, packing fraction, requires knowledge of the cell volume. If one is canvassing new chemical compounds with data mining and requires knowledge of a descriptor that is cell-volume dependent, schemes that can provide accurate estimates of the cell parameters are highly desirable.

In this work, we propose two prediction schemes to provide improved estimates of the lattice lengths (and hence, volume) of a candidate crystal structure. The first scheme, which relies on a one-to-one mapping of species in the candidate crystal structure to a known crystal structure, is able to achieve very low mean absolute errors (MAEs) of 3.8% in the volume, whereas the second scheme, which relies on data-mined minimum atom pair distances, can achieve a MAE of 8.2%. We will also discuss the various factors that may impact the performance of the schemes.

## 2. Lattice length scaling schemes

### 2.1. Reference lattice scaling scheme

In the first scheme, we focus on new materials that are derived from ionic substitutions of a known crystal, i.e., the atomic positions and lattice parameters of the parent structure are known from either experiments or computations. We will henceforth refer to this scheme as the "reference lattice scaling (RLS) scheme".

Assuming that there are no large changes in lattice angles and atomic positions, our hypothesis is that the lengths of the lattice vectors $\{a_i\}$ are proportional to the sum of the atomic density-weighted atomic radii of the species in the crystal structure, as follows,

$$a_i \propto \sum_{k=1}^{3} r_k \cdot (N_k)^{1/3},$$

where $N_k$ and $r_k$ are the number of atoms of specie $k$ in the cell and the atomic radius of specie $k$, respectively, and the factor of 1/3 converts the volume density to a length density. Here, the atomic radii refer to one of the commonly used definitions of ionic, covalent or Van der Waals radii. We will discuss the selection of radii in a later section. We have observed a similar relationship for the case of bournonite (CuPbSbS$_3$) family in our recent work, where the computed cell volume for over 300 substitutions was approximately proportional to the sum of atomic volumes determined by the composition.[18]

The relationship between the lattice lengths of a derived structure $\{a_i^d\}$ can then be related to the parent structure $\{a_i^p\}$ as follows:

$$\frac{a_i^d}{a_i^p} = \frac{\sum_{k=1}^{N} r_k^d \cdot (N_k^d)^{1/3}}{\sum_{k=1}^{N} r_k^p \cdot (N_k^p)^{1/3}} = \alpha_r \qquad (1)$$

where the superscripts $d$ and $p$ are used to label parameters for the derived or parent structures, respectively. Similarly, one can demonstrate that the ratio between the volume of the derived structure $V_d$ and the parent structure $V_p$ is given as follows:

$$\frac{V_d}{V_p} = \left[\frac{\sum_{k=1}^{N} r_k^d \cdot (N_k^d)^{1/3}}{\sum_{k=1}^{N} r_k^p \cdot (N_k^p)^{1/3}}\right]^3 = \alpha_r^3 \quad (2)$$

Figure 1(a) illustrates the schematic application of RLS to a derived structure as an example, in which the initial cell parameters $\{a_i^d\}$ are scaled by the factor $\alpha_r$ defined in Eq. 1.

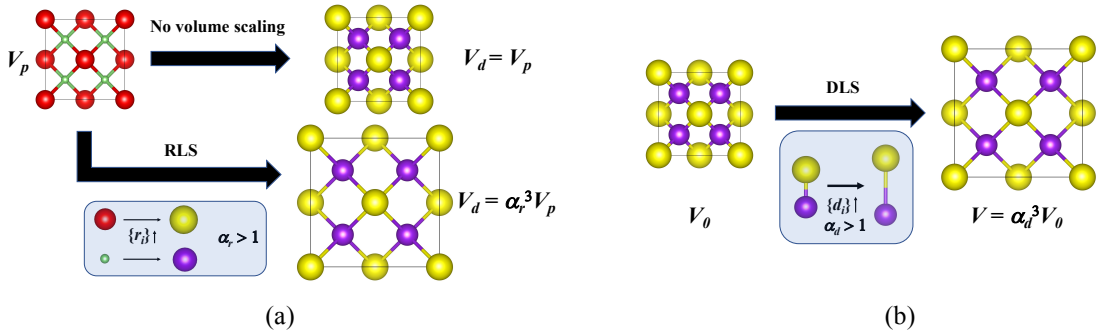

(a) (b)

Figure 1. Schematics of the lattice scaling from (a) reference lattice scaling (RLS) scheme, and (b) data-mined lattice scaling (DLS) scheme. The scaling factors $\alpha_r$ and $\alpha_d$ are defined in Eq. 1 and Eq. 4, respectively.

## 2.2. Data-mined lattice scaling scheme

Unlike RLS, the second scaling scheme for a new material does not require computational or experimental knowledge of a reference crystal. As the predicted crystal parameters are determined based on a data-mining approach, we refer this second scheme as to "data-mined lattice scaling (DLS) scheme".

For a given crystal structure $X$, we scale its lattice parameters by a factor determined based on the data-mined predicted atom pair distance between two atoms in $X$ versus their initial distance. Here, any atoms within 4 Å are considered as potential atom pairs.

In the data-mined predictor, the distance associated with two species $i$ and $j$, $d_{ij}$, is parameterized as

$$d_{ij} = r_i + r_j + \sigma_X k_i + \sigma_X k_j \qquad (3)$$

where $\sigma_X$ is the standard deviation of Pauling electronegativity of all the species in structure $X$, called the "electronegativity spread". The electronegativity spread is intended to be a measure of "structural ionicity": $\sigma_X$ equals zero for any pure element while $\sigma_X$ is large for highly ionic compounds (e.g., $\sigma_X = 1.5$ for LiF). The parameters $r_i$ and $k_i$ are specie dependent, and are derived from fitting $\{(r_i, k_i)\}$ via linear regression on a large training set of observed atom pair distances. In this work, we acquired a large training set of 23,721 thermodynamically-stable (i.e., energy above hull $(E_{\text{hull}})^{2,19,20} = 0$ meV/atom) crystal structures from the Materials Project (MP) database and ran an iterative fitting procedure to determine the $\{(r_i, k_i)\}$ parameters. The fitting procedure and the performance of the DLS on the training set are given in Supplementary Information (SI). We expect the fitted $r_i$ to be approximately equal to the atomic radius ($r_c$) because $r_i$ represents the contribution of an atom to the atom pair distances in the absence of any electronegativity spread ($\sigma_X = 0$), i.e., in a pure element. The fitted $k_i$ is an adjustment factor based on $\sigma_X$ in a material that allows the atomic radius to change in more electronegative compounds to provide a continuous measure of ionic radius and we expect that $k_i$ becomes negative for cations and positive for anions. The associated fitted values are tabulated in Table S1.

After the set of parameter pairs $\{(r_i, k_i)\}$ are trained, the predicted lattice parameters for any input crystal structure with initial lattice parameters $\{a_i\}$ and atom pair distances $\{d_{ij}\}$ can be estimated as $a_i^{DLS} = \alpha_d \cdot a_i$, where the lattice scaling factor $\alpha_d$ is computed from the "most constrained atom distance" as follows,

$$\alpha_d = \max\left\{\frac{d_{ij}^{DLS}}{d_{ij}}\right\} \qquad (4)$$

where $d_{ij}^{DLS}$ is the predicted minimum distance two atoms computed using parameters $\{(r_i, k_i)\}$. Thus, the algorithm simultaneously enforces two conditions: (i) no two atoms are closer than their minimum predicted distance $d_{ij}^{DLS}$, preventing "too small" volumes, and (ii) at least one pair of atoms are at precisely their minimum predicted distance, preventing "too large" volumes. Figure 1(b) depicts the schematics of DLS.

## 3. Selection of test set

To evaluate the performance of our proposed schemes, we selected a test set of 3112 structure pairs ($S_p$, $S_d$) from 309 structural prototypes in the 2016 version of Inorganic Crystal Structure Database (ICSD).[21] $S_p$ and $S_d$ refer to the parent and derived structures, respectively. All selected derived structures in the test set are subject to the following constraints.

(i) They are unique by the fact that each of them is mapped to the parent structure with the lowest electronegativity ($\chi$) difference.

(ii) They have $E_{hull}$ less than 50 meV/atom. This is to avoid the inclusion of unstable phases, some of which are synthesized under high pressure conditions, within the test set.

(iii) They have an associated DFT-PBE computed volume available in the MP database.

(iv) They do not contain noble gas species.

(v) Only like-charge substitutions are allowed, i.e., no substitutions of cations with anions are allowed.

It should be noted that for DLS, only the derived structures $\{S_d\}$ are used in the evaluation because it does not require knowledge of the ionic substitutions.

We also evaluate the performance of both lattice scaling schemes on compounds that are thermodynamically stable. For this, we select a subset of the test set such that all parent and derived structures in the subset have $E_{\text{hull}} = 0$ meV/atom, creating a set of 2,129 derived structures. We denote this subset as "stable test set" in the following.

The distribution of species and number of species in the test set are presented in Figure 2. We find that the majority are ternary and quaternary compounds (see Figure 2(a)), and they have a reasonably good coverage of elements in the periodic table, with half of the derived structures containing oxygen (Figure 2(b)).

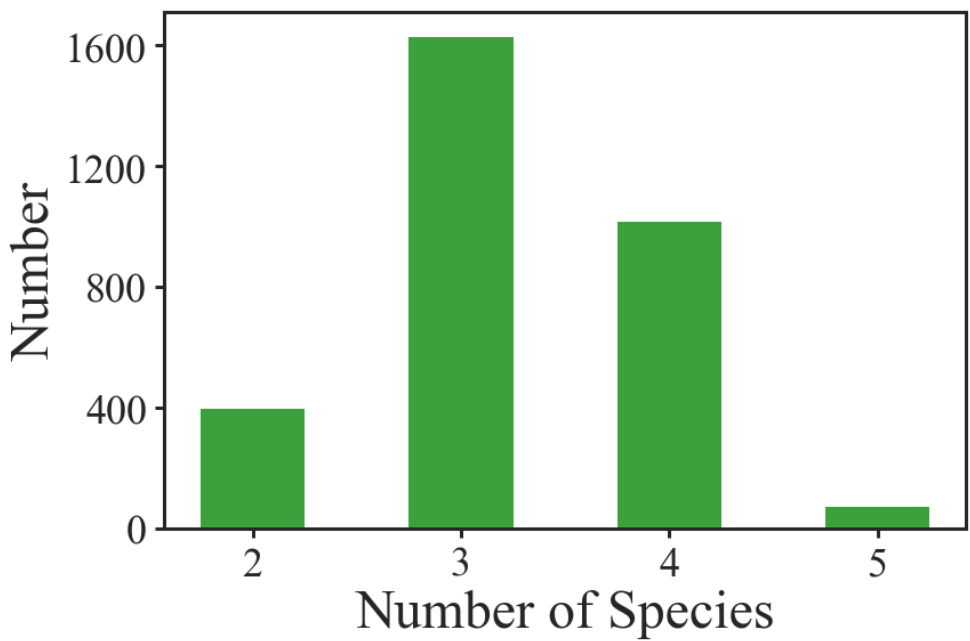

(a)

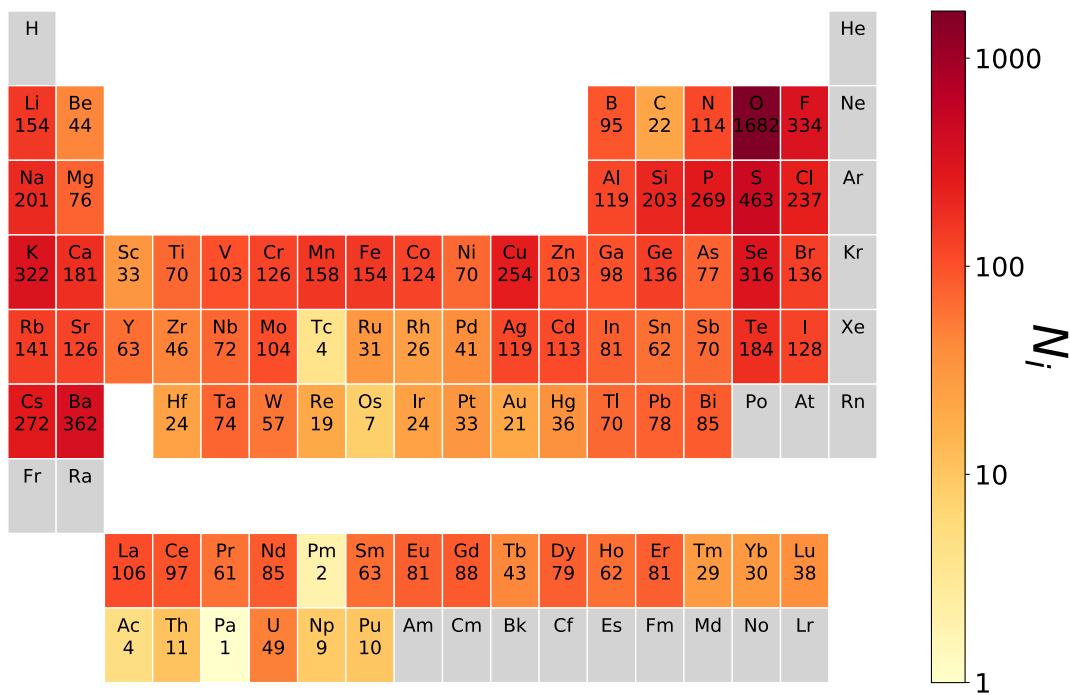

(b)

Figure 2. (a) Distribution of appearing species in the *derived* structures of the entire test set. (b) Color map indicates frequency of each element appearing in all derived structures of the test set in a log scale, with red (yellow) corresponding to a high (low) value. The corresponding values for the elements are also provided. Elements with zero value are highlighted in grey.

## 4. Performance evaluation

In this section, we present the performance of both lattice scaling schemes using the test set. For comparison, we also estimate the volume prediction error of a commonly used procedure, in which the cell volume of the derived structure is simply set as that of the parent structure (see also Figure 1(a)). We refer this scheme to as "unscaled reference lattice scheme".

### 4.1. Reference lattice scaling scheme (RLS)

The DFT-PBE volume prediction errors of RLS using ionic radii and covalent radii are evaluated. Here, the volume prediction error is the percentage error of the predicted volume compared to the actual value. Figure 3(a) and (b) (Table 1) shows the histogram of prediction errors for the entire test set ($E_{hull} \leq 50$ meV/atom). Similarly, Figure 4(a) and (b) (Table 2) present the same results for the stable test set ($E_{hull} = 0$ meV/atom). We find that RLS using ionic radii leads to the lowest MAE of 3.8% for the prediction error, lower than that using covalent radii (4.9%). For the unscaled reference lattice scheme, however, the MAE is significantly higher (9.3%). We also note that the distribution of prediction error depends only weakly on the selected $E_{hull}$ threshold.

The ICSD experimental volume prediction error with RLS using the same training set was also evaluated. The results are given in Figure S3(a) and S3(b) (Table S4) for the entire test set ($E_{hull} \leq 50$ meV/atom), and Figure S4(a) and S4(b) (Table S5) for the stable test set ($E_{hull} = 0$ meV/atom) in SI. Once again, RLS using ionic radii has the lowest MAE of 4.3%, lower than that using covalent radii (5.7%) and that without lattice scaling (9.0%). This confirms the general applicability of the proposed RLS scheme to both ICSD experimental volumes as well as DFT relaxed volumes.

We note that the relatively small prediction error of the unscaled reference lattice scheme (<10%) is mainly due to the fact that we used the structure with the smallest mean absolute electronegativity difference as the reference. As an example, in calculating the error in the predicted volume of LiF, the volume of NaF is used instead of that from KF. Given that elements with similar electronegativities tend to have similar radii, it is therefore not surprising that the volumes from the unscaled reference lattice scheme are relatively good estimates. Nevertheless, the RLS scheme still outperforms the unscaled scheme by more than a factor of 2. To probe the effect of choice of the reference structure, we performed the same analysis using structure pairs with the maximum mean absolute electronegativity difference as an evaluation of the performance of the RLS under the worst-case scenario. The resulting MAE for the test set ($E_{hull}$ ≤ 50 meV/atom) using DFT-PBE volumes are 8.4%, 15.1% and 30.2% for RLS using ionic radii, RLS with covalent radii, and unscaled reference lattice scheme, respectively (see Table 1). The associated prediction error distributions are provided in SI (see Figure S5). In other words, the RLS scheme outperforms the unscaled scheme by an even greater margin if non-ideal reference structures (as defined by electronegativity difference) are used.

Unless otherwise specified, we will henceforth discuss only the results of the RLS scheme using the ionic radii and the test set with the minimum mean absolute electronegativity difference. We should note that if the ionic radii of some species are not available, the covalent radii can be used as an effective fallback for the RLS scheme.

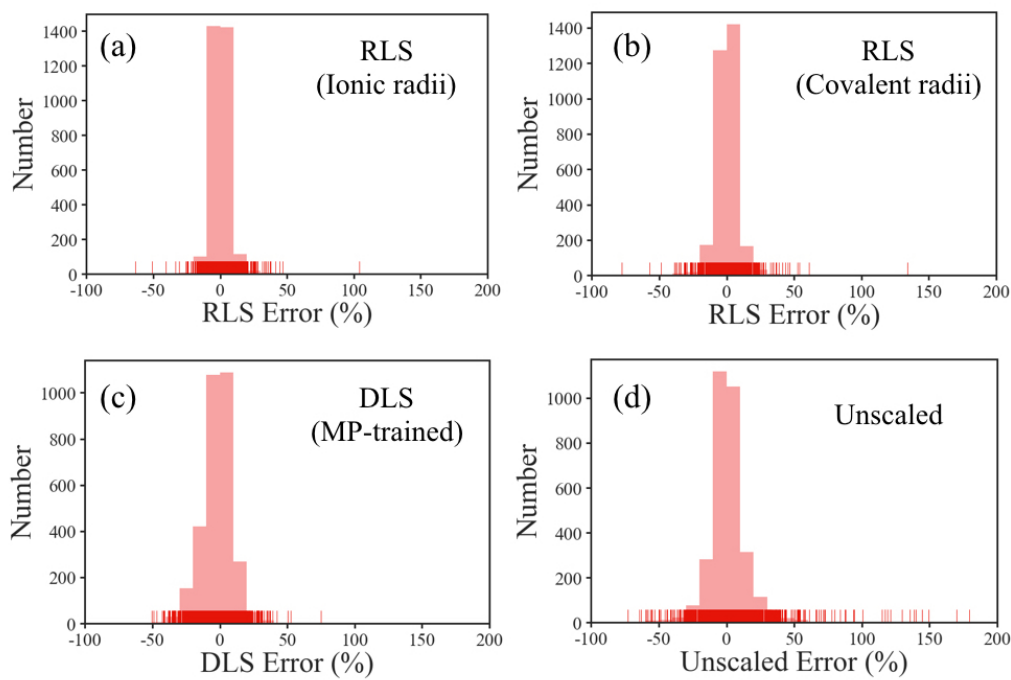

Figure 3. Histograms of the DFT-PBE volume prediction error of the entire test set ($E_{hull}$ ≤ 50 meV/atom, minimum electronegativity difference criterion for reference structures) using (a) RLS with ionic radii, (b) RLS with covalent

radii, (c) DLS, and (d) unscaled reference lattice scheme. Key variables of the error distribution are also listed in Table 1.

Table 1. Key variables that describe the histogram of DFT-PBE volume errors of the test sets selected using (i) minimum and (ii) maximum electronegativity difference criteria in the cases of RLS, DLS, and unscaled reference lattice schemes. In these test sets, all structures have $E_{hull} \leq 50$ meV/atom.

| Reference selection | Scheme | MAE(%) | $\sigma$ (%) | Max. error (%) | Min. error (%) |
|---|---|---|---|---|---|
| Minimum electronegativity difference | RLS + ionic radii | 3.8 | 6.5 | 104 | -63.3 |
| | RLS + covalent radii | 4.9 | 8.1 | 134 | -77.8 |
| | DLS | 8.2 | 11.1 | 74.9 | -50.7 |
| | Unscaled | 9.3 | 16.4 | 179 | -72.7 |
| Maximum electronegativity difference | RLS + ionic radii | 8.4 | 13.1 | 200 | -66.7 |
| | RLS + covalent radii | 15.1 | 23.8 | 423 | -80.7 |
| | DLS | 8.2 | 11.1 | 74.9 | -50.7 |
| | Unscaled | 30.2 | 49.1 | 452 | -81.9 |

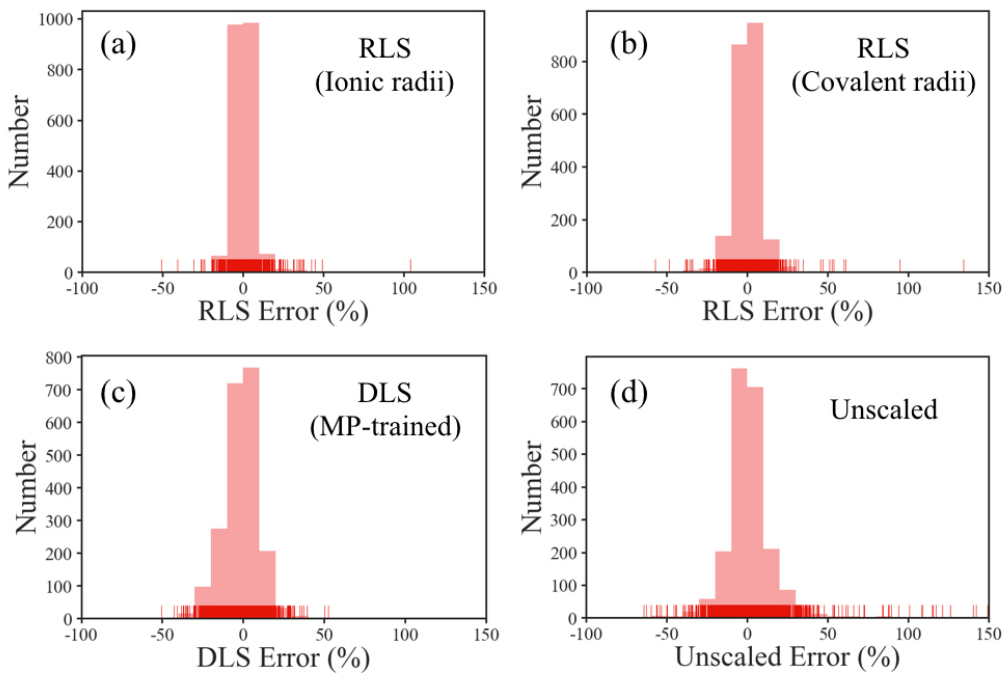

Figure 4. Histograms of the DFT-PBE volume prediction error of the stable test set ($E_{hull}$ = 0 meV/atom, minimum electronegativity difference criterion for reference structures) using (a) RLS with ionic radii, (b) RLS with covalent radii, (c) DLS, and (d) unscaled reference lattice scheme. Key variables of the error distribution are also listed in Table 2.

Table 2. Key variables that describe the histogram of DFT-PBE volume errors of the test sets selected using (i) minimum and (ii) maximum electronegativity difference criteria in the cases of RLS, DLS, and unscaled reference lattice schemes. In these test sets, all structures have $E_{\text{hull}} = 0$ meV/atom.

| Reference selection | Scheme | MAE(%) | $\sigma$ (%) | Max. error (%) | Min. error (%) |
|---|---|---|---|---|---|
| Minimum electronegativity difference | RLS + ionic radii | 3.9 | 6.7 | 104 | -51.0 |
| | RLS + covalent radii | 5.3 | 8.8 | 134 | -57.3 |
| | DLS | 8.0 | 10.8 | 52.5 | -50.7 |
| | Unscaled | 9.7 | 17.2 | 179 | -64.2 |
| Maximum electronegativity difference | RLS + ionic radii | 7.9 | 11.9 | 83.7 | -49.2 |
| | RLS + covalent radii | 14.1 | 20.6 | 185 | -65.0 |
| | DLS | 8.0 | 10.8 | 52.5 | -50.7 |
| | Unscaled | 27.9 | 45.4 | 452 | -81.9 |

Figure 5 plots the RLS prediction error of the entire test set vs. (a) composition-weighted average ionic radii difference $\overline{\Delta r}$, and (b) ionic volume ratio difference ($\Delta\eta = |\eta_d - \eta_p|$, where $\eta = [\frac{4\pi}{3}\sum_i r_i^3]/V$) between the derived and parent structures. Here, DFT-PBE volumes are used in the analysis. Overall, we observe from these plots that the distribution of the structure pairs is concentrated in a narrow region (indicated as red) where both the prediction error and the factors studied, i.e., ionic radii difference and ionic volume ratio difference, are small.

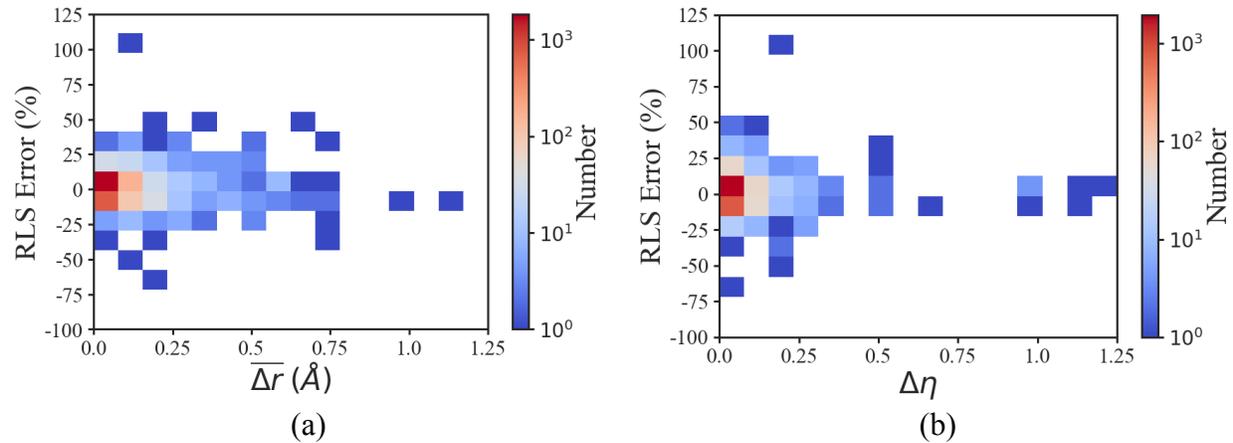

Figure 5. (a) RLS DFT-PBE volume error as a function of (a) composition-weighted average ionic radii difference $\overline{\Delta r}$, and (b) ionic volume ratio difference ($\Delta\eta$) for the entire test set with minimum electronegativity difference.

Here, color map is used to indicate the number of data points within each given region, with red (blue) corresponding to a high (low) value. No points fall in the white region.

### 4.1.1. Analysis of outliers

When DFT-PBE volumes are used, we find that there are three outliers, i.e., structure pairs where the absolute volume prediction error is > 50%: (i) $BiF_3$ (mp-23237, *Pnma*) from $AsF_3$ (mp-28027, $Pna2_1$), with prediction error ~ 100%; (ii) $AsF_3$ (mp-28027, $Pna2_1$) from $BiF_3$ (mp-23237, *Pnma*), with prediction error ~ -51%; (iii) $CO_2$ (mp-556034, *Pbcn*) from $SnO_2$ (mp-12978, *Pbcn*), with prediction error ~ -63%. In all these cases, either the parent or derived structure is a molecular crystal. When ICSD volumes are used, there are three additional outliers. One of them is BeO (mp-1778, $F\bar{4}3m$) from CoO (mp-24864, $F\bar{4}3m$), in which we find that the experimental volume of BeO is from high-pressure synthesis. For the other two outliers, the derived structures are obtained via aliovalent substitution, and the corresponding volume prediction errors are slightly above 50%.

### 4.2. Data-mined lattice scaling scheme (DLS)

The MAE of DLS error of the entire test set using DFT-PBE volumes is 8.2%. We note that this error is very similar to the computed MAE for the training set (8.1%, see SI), which was obtained using the $\{\{r_i, k_i\}\}$ fitting procedure on a wider spectrum of compounds in the MP database. Figure 3(c) (Table 1) depicts the performance for the entire test set ($E_{hull} \leq 50$ meV/atom), while Figure 4(c) (Table 2) presents results for the stable test set ($E_{hull} = 0$ meV/atom). Overall, the MAE of the volume error is reduced when the DLS scheme is applied (8.2%) compared to the unscaled reference lattice scheme (9.3%). We note that the distribution of prediction error with $E_{hull} \leq 50$ meV/atom is very similar to that with $E_{hull} = 0$ meV/atom, confirming that the DLS scheme generalizes well to metastable compounds.

The advantage of using DLS is magnified when the test set is obtained with the maximum electronegativity difference. In this case, the 8.2% MAE of DLS vastly outperforms that of the unscaled lattice reference scheme (30.2%) and is even comparable to that of the RLS method with ionic radii (8.4%) despite not having prior information about a reference structure.

We also tested the performance of DLS in predicting ICSD volumes rather than DFT-PBE volumes. Strictly speaking, one should refit the $\{\{r_i, k_i\}\}$ parameters for this situation. However, we verify the transferability of existing parameters by instead simply including an additional scaling factor of 1.05 that accounts for the fact that DFT-PBE tends to result in lattice volumes that are ~5% larger than experiments.[22] In this case, the MAE of DLS error for the entire test set ($E_{hull} \leq 50$ meV/atom) is 9.7% (see Figure S3(c) and Table S4 in SI), while the MAE for the stable test set ($E_{hull} = 0$ meV/atom) is 8.8%, as shown in Figure S4(c) and Table S5 in SI. Although the MAE of DLS prediction error is similar to that of the unscaled reference lattice scheme (with knowledge of a "good" reference), the number of outlier cases are greatly reduced, as observed in Figure S3 and Figure S4.

Figure 6 plots the DFT-PBE volume prediction error of DLS vs. the standard deviation of Pauling electronegativity in a given structure $X$ ($\sigma_X$, see Figure 6(a) for $E_{hull} \leq 50$ meV/atom; and Figure 6(b) for $E_{hull} = 0$ meV/atom). Overall, the plots reveal that there is no major correlation between the DLS error and the electronegativity spread. In addition, there is no clear trend between the prediction volume error and number of atoms (Figure S6) and/or number of species (Figure S7).

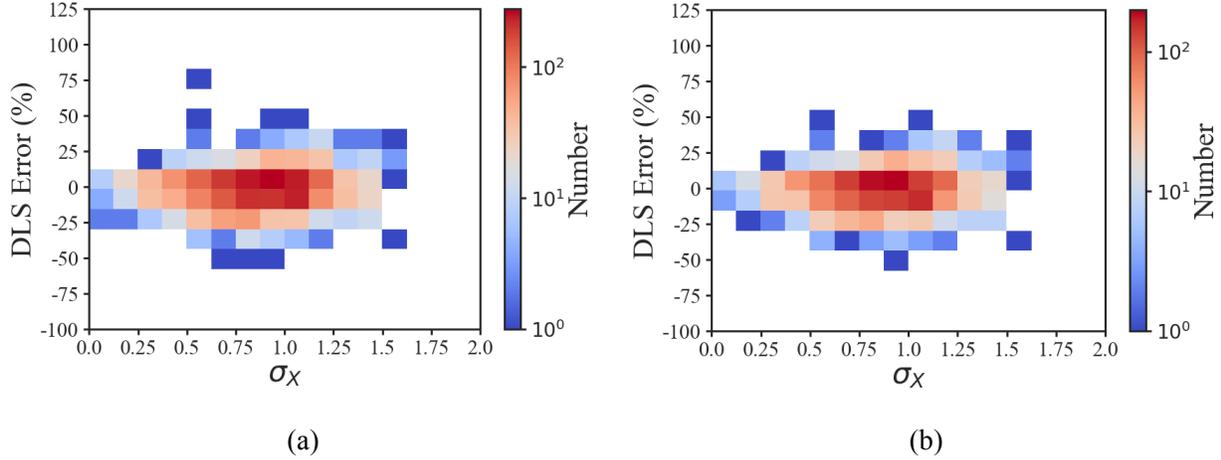

Figure 6. DLS DFT-PBE volume prediction error as a function of the standard deviation of Pauling electronegativity $\sigma_X$ for test set with (a) $E_{hull} \leq 50$ meV/atom and (b) $E_{hull} = 0$ meV/atom. Here, color map is used to indicate the number of data points within each given region, with red (blue) corresponding to a high (low) value. No points fall in the white region.

### 4.2.1. Analysis of outliers

There are 4 outliers (absolute error > 50%) when DFT-PBE volumes are used: (i) $CsV_5S_8$ (mp-985699, $C2/m$), (ii) $CsCr_5S_8$ (mp-540569, $C2/m$), (iii) $CsMnO_4$ (mp-18994, $Pnma$), (iv) $HgF_2$ (mp-8177, $Fm\bar{3}m$). Outliers (i) and (ii) occur because their volume scaling is based on the Cs-Cs which is one of the poorest predicted atom pair distances using this model. For (iii), the volume prediction for $CsMnO_4$ is based off the Mn-O atom pair distance, which overestimates the atom pair distance at approximately 1.85 Å while the actual distance is approximately 1.6 Å – the mean distance of Mn-O in the database is 1.85 Å. This is likely due to an unusual formal oxidation state of $Mn^{6+}$ in this compound. For (iv), the volume prediction for $HgF_2$ is based on the F-Hg atom pair distance, estimated at 1.93 Å vs actual distance of 2.44 Å.

There are an additional 25 outliers when ICSD volumes are used in conjunction with the additional scaling factor (1.05). Of these, 23 of the ICSD predicted volumes have a greater than 25% difference compared to the DFT-PBE prediction (19 are greater than 50% difference). The two other outliers are (i) $CsCr_5S_8$ and (ii) $HgF_2$, which are explained above.

## 4.3. Error comparison between RLS and DLS on the test set

Figure 7 compares the volume prediction error of RLS versus that of DLS, in which the outliers presented in the previous section, i.e. those with prediction error > 50%, are highlighted. We find that there are no common outliers for both RLS and DLS schemes. Moreover, our results suggest that DLS tends to outperform RLS for the case of molecular crystals, e.g. $AsF_3$ (mp-28027, $Pna2_1$) and $CO_2$ (mp-556034, $Pbcn$), which are the major outliers identified for the latter scheme. However, for compounds containing species with multiple oxidation states, e.g. Mn, DLS can lead to much higher prediction error than that of RLS, e.g., $CsMnO_4$ (mp-18994, $Pnma$).

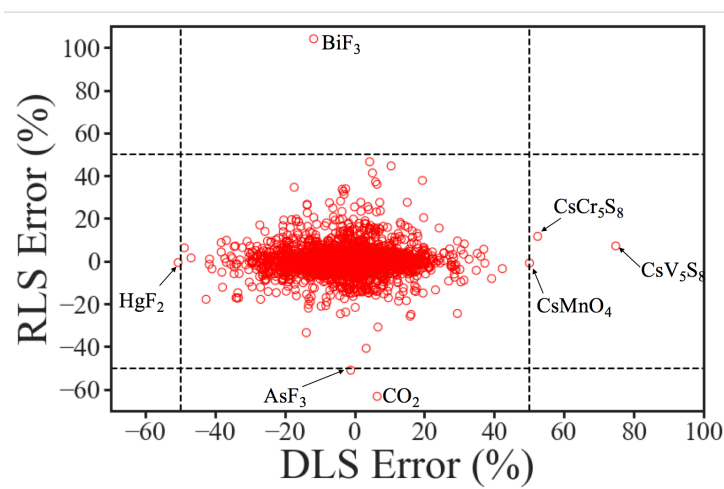

Figure 7. Comparison of prediction errors between DLS and RLS using the DFT-PBE volumes of the entire test set. Dotted lines mark the 50% error cutoff. Outliers with absolute error greater than 50% for each prediction scheme are labeled.

## 4.4. Performance in DFT structural optimizations

Given that RLS using ionic radii outperforms other lattice scaling schemes, we here estimate its effect on convergence in DFT structural optimizations. The computational details are provided in SI for interested readers. We selected one structure pair per prototype from the test set that has at least 25 atoms per unit cell. For each structure pair ($S_A$, $S_B$), we generate two derived structures, (i) $S_A^*$ from $S_B$, and (ii) $S_B^*$ from $S_A$. As the optimization approach also plays an important role in the convergence speed of DFT relaxations, we compare the performance of two widely used optimization approaches: (i) the conjugate gradient (CG) approach, and (ii) the quasi-Newton (QN) approach. For each derived structure, the DFT relaxation is performed under four conditions, (i) RLS + QN approach, (ii) RLS + CG approach, (iii) unscaled reference lattice scheme + QN approach, and (iv) unscaled reference lattice scheme + CG approach. In the end, there are 169 derived structures in which DFT relaxation is properly converged under all four conditions.

We first estimated the initial volume error percentage ($\Delta V_{err}$) of RLS with respect to unscaled reference lattice scheme, $\Delta V_{err} = (|V_{unscaled} - V_{true}| - |V_{RLS} - V_{true}|) / V_{true}$, where a positive value indicates a smaller initial volume error by RLS than that without lattice scaling. Figure 8(a) plots the distribution of $\Delta V_{err}$ of the selected 169 derived structures. 110 out of them have positive $\Delta V_{err}$. This ratio (~65%) is slightly lower to that when the entire test set is considered, in which 2321 out of 3112 structures (~75%) have positive $\Delta V_{err}$. This suggests that RLS generally improves the initial volume of the derived structures.

We then estimated the potential speedup using RLS in terms of the number of total electronic steps ($N_e$) in the DFT relaxation. Specifically, we compare the performance of RLS vs. that of unscaled reference lattice scheme using the same optimization approach. When QN is adopted, we find that 88 out of 169 calculations exhibit speedup upon using RLS compared to those without lattice scaling. For CG, 79 out of 169 calculations exhibit speedup using RLS.

Figure 8(b) plots $N_e$ of unscaled reference lattice scheme vs. $N_e$ of RLS using these two optimization schemes, in which only the large-$N_e$ region is depicted. In the figure, data points above the line with slope $k = 1$ suggests speedup upon using RLS, whereas those below the line with $k = 1/2$ suggests the $N_e$ of RLS is at least a factor of two greater than that of unscaled reference lattice scheme, i.e., RLS slows down the convergence of DFT relaxations. There are two data points that are below the line with $k = 1/2$ (see Figure 8(b)), (i) KGe$_2$(PO$_4$)$_3$ (mp-18203, $R\bar{3}$) from LiGe$_2$(PO$_4$)$_3$ (mp-541272, $R\bar{3}c$) ($N_e$ ratio ~ 2.23; $\Delta V_{err}$ ~ -30.4%), and (ii) BaBSbS$_4$ (mp-866301, $Pnma$) from KBaNbS$_4$ (mp-16780, $Pnma$) ($N_e$ ratio ~ 2.42; $\Delta V_{err}$ ~ -17.1%). Both cases have negative $\Delta V_{err}$, suggesting the initial volume error upon RLS is larger than that without lattice scaling.

We also compared between the two optimization schemes (QN or CG) when RLS is applied. We find that RLS + QN performs better than RLS + CG in 149 out of 169 calculations. Figure 8(c) plots $N_e$ of RLS + CG vs. that of RLS + QN. There are two cases where $N_e$ using QN is a factor of two larger than that using CG (the points fall below the line with $k = 1/2$), (i) BaBSbS$_4$ (mp-866301, $Pnma$) from KBaNbS$_4$ (mp-16780, $Pnma$) ($N_e$ ratio ~ 2.12; $\Delta V_{err}$ ~ -17.1%), and (ii) CaCu(GeO$_3$)$_2$ (mp-6537, $P2_1/c$) from LiFe(GeO$_3$)$_2$ (mp-645305, $P2_1/c$) ($N_e$ ratio ~ 2.14; $\Delta V_{err}$ ~ -9.3%). Once again, both cases have negative $\Delta V_{err}$, suggesting that the initial guess of the cell volume is worse than that without volume scaling.

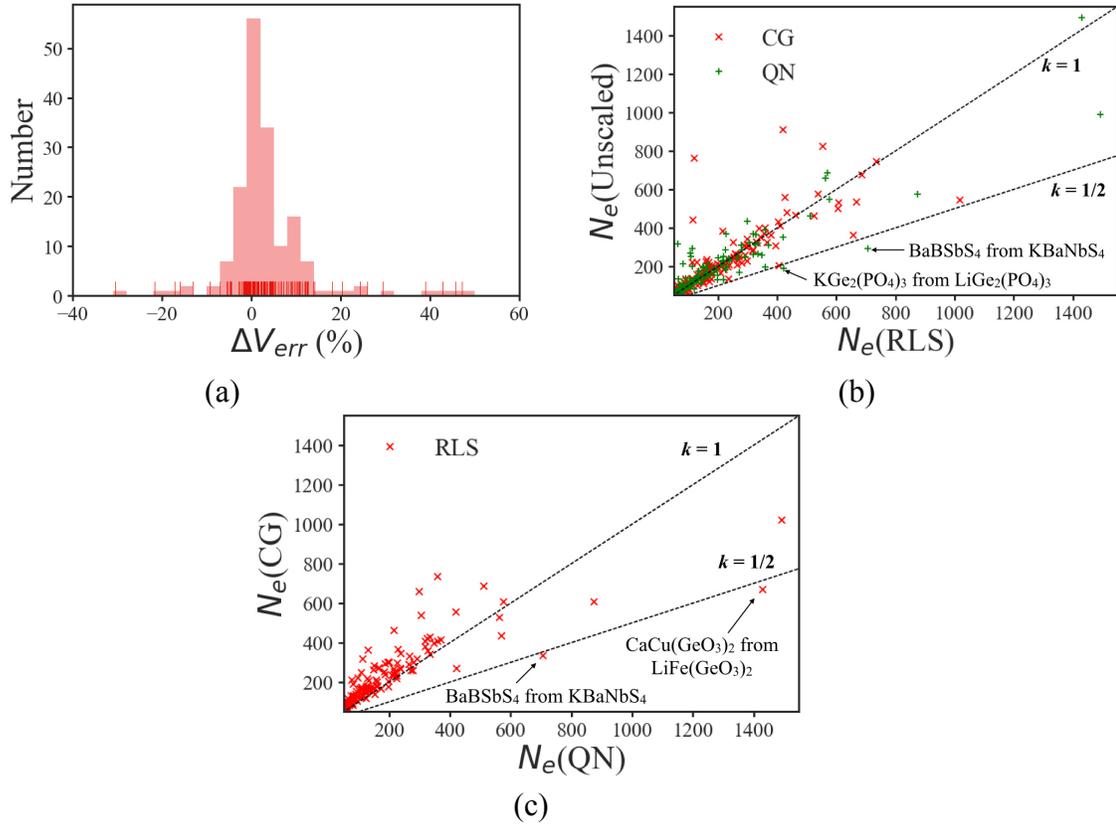

Figure 8. (a) Distribution of the initial volume error percentage ($\Delta V_{err}$), where a positive value indicates a smaller initial volume error by RLS than that of unscaled reference lattice scheme. (b) $N_e$ of unscaled reference lattice scheme versus that of RLS. Here, red (green) cross markers correspond to the geometry optimizations using the CG (QN) approach. Dashed lines with given slope values $k=1$ and $1/2$ are provided to help identifying the outliers, and substitutions falling below the line with $k=1/2$ are labeled (see the text for details). (c) Comparison of $N_e$ using CG and QN optimization approaches when RLS is applied. Substitutions with high $N_e(QN)/N_e(CG)$ ratio are labeled (see the text for details).

## 5. Discussion

Substitution of various species in a known crystal (reference structure) is a common strategy for generating new structures. In this work, we propose two lattice scaling schemes to improve such initial guess for the structural optimization: (1) reference lattice scaling (RLS) that requires knowledge of a reference structure, and we demonstrate that usage of ionic radii leads to the best performance compared to the case of covalent radii; (2) data-mined lattice scaling (DLS) in which the knowledge of reference structure is not needed.

There exist some common limitations for both the RLS and DLS schemes proposed in this work. First, both schemes assume isotropic lattice scaling. For candidate structures that have anisotropic atomic arrangement, e.g., lithium layered transition metal oxides such as $LiCoO_2$ and $LiNiO_2$, the two schemes may lead to additional prediction error if the anisotropy is not accurately reflected in the original / reference structure. Second, we note that while the RLS algorithm does consistently lead to better predictions of volume than an "unscaled" scheme, its effect on DFT convergence speed is much less pronounced. The speed of the convergence not only depends on the provided initial guess of lattice parameters and atomic positions, but is also affected by the structural optimization approach used, e.g., CG or QN approach, and the optimization parameters. It should be noted though that the "unscaled" scheme still assumes that a derived structure is obtained from the minimum absolute electronegativity difference substitution from a list of crystals with the same prototype, which tend to minimize the error in volumes even without scaling. Under the more common situation where a new crystal is derived from any prototype available to the researcher, we expect the volume error of the "unscaled" scheme to be larger on average.

For RLS, we should note that the scaling factor given in Eq. 1 does not consider the effect of atomic packing. Specifically, for two structures with the same composition but different atomic packings, the scaling factor is identical for both structures per Eq. 1. Moreover, when substitution between cations and anions occurs, RLS scheme can yield high prediction error because the strong local structural distortion is not considered in the lattice scaling.

For DLS, the proposed scheme does not consider the oxidation state when estimating the distance between two atoms, which could explain the high standard deviation of certain atom pairs. Note that the DLS model could in theory treat different ions as different species with no further modifications to the formalism (but would require refitting the new parameters). However, this would require a careful tagging of ions in the training data as well as knowledge of oxidation state for new compounds used in prediction. With the current scheme, knowledge of oxidation state is not needed to perform a volume prediction. Figure S8 in SI presents a box and whisker plot that displays the 15 atom pairs that have a standard deviation greater than 0.2 Å in the training set from MP database. Such atom pairs that exhibit high variance in atomic distance that may be due to different oxidation states (including the metallic state) that are being averaged into the same species.

Finally, there are a few scenarios under which the DLS scheme can be applied while RLS may not be applicable. (i) When the parent structure is not known, i.e., for completely new structural prototypes, or when one cannot match a candidate structure to known prototypes using a cation→cation and anion→anion matching. (ii) When molecular crystals are concerned, DLS tends to outperform RLS scheme. (iii) When not all the ionic radii are present for the candidate structure, RLS must utilize covalent radii as fallback while DLS does not require knowledge of atomic radii.

## 6. Code availability

Both RLS and DLS schemes are implemented in Python Materials Genomics (pymatgen), an open-source Python library for materials analysis. An example script for the usage of both schemes are provided in SI.

## 7. Conclusion

To summarize, we propose two lattice scaling schemes that improve estimates of the lattice parameters of a candidate structure. The first scheme is reference lattice scaling (RLS) that requires knowledge of reference crystal structure, while the second scheme is data-mined lattice scaling (DLS) that employs data-mined minimum atom pair distances to predict the crystal volume of the candidate structure. We demonstrate that both RLS and DLS can effectively predict the crystal volume with a mean absolute error as low as 3.8% and 8.2%, respectively.

## Acknowledgements

I.-H. Chu, D. Han and S. P. Ong were supported by the Materials Project, funded by the U.S. Department of Energy, Office of Science, Office of Basic Energy Sciences, Materials Sciences and Engineering Division under Contract No. DE-AC02-05-CH11231: Materials Project program KC23MP, which intellectually led this work. S. Roychowdhury and A. Jain were supported by the U.S. Department of Energy, Office of Basic Energy Sciences, Early Career Research Program. Additional funding for S. Roychowdhury was provided by the U.S. Department of Energy, Office of Science, Office of Workforce Development for Teachers and Scientists (WDTS) under the Science Undergraduate Laboratory Internship (SULI) program. We also acknowledge computational resources provided by Triton Shared Computing Cluster (TSCC) at the University of California, San Diego, the National Energy Research Scientific Computing Center (NERSC, a DOE Office of Science User Facility supported by the Office of Science of the U.S. Department of Energy under Contract No. DE-AC02-05CH11231), and the Extreme Science and Engineering Discovery Environment (XSEDE) supported by National Science Foundation under Grant No. ACI-1053575.

## Appendix A. Supplementary material

Supplementary data associated with this article can be found in the online version.